\documentclass[aps,prl,twocolumn,floatfix,showpacs,showkeys]{revtex4}

\usepackage{graphicx}
\usepackage{dcolumn}
\usepackage{bm}

\begin{document}


\title{Spontaneous supercurrent induced by ferromagnetic pi-junctions}

\author{A. Bauer$^1$, J. Bentner$^1$, M. Aprili$^{2}$, M. L. Della Rocca$^{2}$, M. Reinwald$^1$, W. Wegscheider$^1$, and C. Strunk$^{1}$}
\email{
christoph.strunk@physik.uni-regensburg.de}
\affiliation{
$^{1}$Institut f\"ur Experimentelle und Angewandte Physik, Universit\"at Regensburg, 93040 Regensburg, Germany\\%
$^{2}$CSNSM-CNRS, Universit\'{e} Paris-Sud, 91405 Orsay Cedex, France}

\altaffiliation[Also at ]{
LPQ-ESPCI, 75005 Paris, France}

\date{\today}

\begin{abstract}
We present magnetization measurements of mesoscopic superconducting niobium loops containing a ferromagnetic (PdNi) pi-junction. The loops are prepared on top of the active area of a micro Hall-sensor based on high mobility GaAs/AlGaAs heterostructures. We observe asymmetric switching of the loop between different magnetization states when reversing the sweep direction of the magnetic field. This provides evidence for a spontaneous current induced by the intrinsic phase shift of the pi-junction. In addition, the presence of the spontaneous current near zero applied field is directly revealed by an increase of the magnetic moment with decreasing temperature, which results in half integer flux quantization in the loop at low temperatures.
\end{abstract}


\maketitle

As first predicted by Josephson\cite{Josephson1962}, Cooper pairs can be transferred coherently between two weakly coupled superconducting electrodes at zero bias voltage. The resulting supercurrent through such a weak link is driven by the phase difference $\varphi$ of the superconducting wavefunction across the junction. In conventional tunnel coupled Josephson junctions the current-phase relation, which connects the supercurrent $I_S$ flowing through a junction and the phase difference of the superconducting wave function or pair amplitude, is given by $I_S(\varphi)=I_C\sin(\varphi)$\cite{Josephson1964}. The critical current $I_C$ is the maximum supercurrent the junction is able to sustain. The sinusoidal shape of the current-phase relation is expected to be modified if the weak link is realized by a ballistic point contact (ScS), or a diffusive normal metal bridge (SNS)\cite{Kulik1978,Heikkilae2002} rather than by a tunneling barrier.

\begin{figure}
\includegraphics{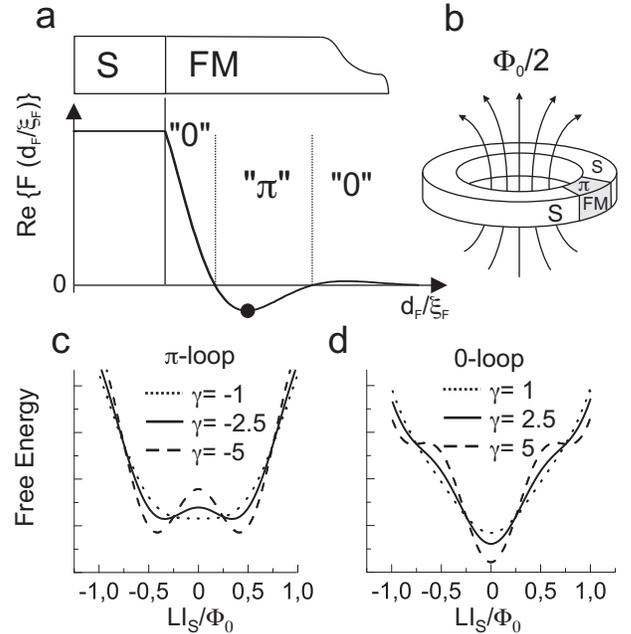}
\caption{\label{fig1} 
The origin and signature of $\pi$-junctions with ferromagnetic weak links. a) Inside the ferromagnet (FM) the real part of the Cooper pair wave function F exhibits spatial oscillations with wavelength $\xi_F$. The dot marks the value of  $d_F / \xi_F$ used in our $\pi$-junctions. \label{fig1b}b) In zero external magnetic field, a spontaneous flux $\Phi = \Phi_0 / 2$ is required to maintain the condition of fluxoid quantization if a $\pi$-junction is inserted into a superconducting loop. c,d) Sketches of the free energy of a $\pi$-loop \label{fig1c}(c) and a 0-loop \label{fig1d}(d) as a function of the magnetic flux for different values of $\gamma \sim I_C$ as described in the text.}
\end{figure}

If the weak link consists of a thin ferromagnetic layer (SFS), the result can be a Josephson-junction with a built-in phase difference of $\pi$. This has been predicted by Buzdin et al.\cite{Buzdin1982}, who calculated that the pair amplitude $F$ inside a ferromagnet in contact with a superconductor should oscillate as a function of position (see Fig. \ref{fig1}a). The origin of the oscillation is the exchange splitting of the conduction bands in the ferromagnet, in which the Cooper pairs acquire a nonzero total momentum. The oscillation period $\xi_F$ is determined by the exchange field. Thus, by choosing appropriate values of exchange field and the thickness $d_F$ of the ferromagnetic layer the ratio $F1/F2$ of the pair amplitude on both sides of the junction can become negative, and with it the current through the junction\cite{Buzdin1982}. Such junctions are called $\pi$-junctions, because the difference in sign between $F_1$ and $F_2$ corresponds to a phase difference of $\pi$ across the junction and hence $I_S(\varphi)=I_C \sin(\varphi + \pi)$. It was pointed out by Bulaevskii et al.\cite{Bulaevskii1977}, that a superconducting ring with an inserted $\pi$-junction exhibits a spontaneous current and a corresponding magnetic flux of \itshape{half}\upshape\/ a flux quantum $\Phi_0$ in the \itshape{ground state}\upshape\/ (see Fig. \ref{fig1b}b). The spontaneous supercurrent has to flow even in absence of external flux in order to maintain the uniqueness of the wave function along the loop provided that the critical current is sufficiently large. This is illustrated in Fig.\ref{fig1}c, where we have plotted the free energy of the $\pi$-loop in absence of an applied field\cite{Bulaevskii1977} as a function of the normalized flux through the loop $LI_S / \Phi_0$ for different values of $\gamma = \frac{1}{\pi \Phi_0} LI_C(T, \frac{d_F}{\xi_F})$, where L is the loop inductance. Here we assume a sinusoidal $I_S(\varphi)$ for simplicity\cite{Kulik1978}. For $\gamma = 1$ a phase transition to a ground state with two degenerate minima corresponding to half a flux quantum occurs in the loop. In Fig. \ref{fig1}d we have plotted the free energy for a 0-loop for comparison.

The first experimental indication for the oscillatory character of the induced superconductivity in thin ferromagnetic films was the observation of a non-monotonic suppression of $T_C$ with increasing film thickness of the ferromagnet in SF multilayers (see e.g.\cite{Chien1999}). An important step towards $\pi$-coupling in magnetic Josephson junctions was the introduction of diluted ferromagnets (e.g. $Cu_xNi_{1-x}$\cite{Ryazanov2001} or $Pd_xNi_{1-x}$\cite{Kontos2001}), which allow larger layer thicknesses for $\pi$-junctions than concentrated ferromagnets. Recently several experiments have revealed more direct evidence of $\pi$-coupling in planar ferromagnetic Josephson junctions\cite{Ryazanov2001,Kontos2001,Kontos2002} and in non-equilibrium superconductor/normal-metal junctions\cite{Baselmans1999,Baselmans2002_1,Baselmans2002_2}. In addition, first phase sensitive measurements on suitably oriented grain boundary junctions with high temperature superconductors\cite{Kirtley1995} as well as 0-$\pi$ dc-SQUIDs\cite{Guichard2003} and networks of $\pi$-junctions\cite{Ryazanov2002} were performed to obtain a direct signature of the $\pi$-shift of the phase of the superconducting wave function. 

\begin{figure}
\includegraphics{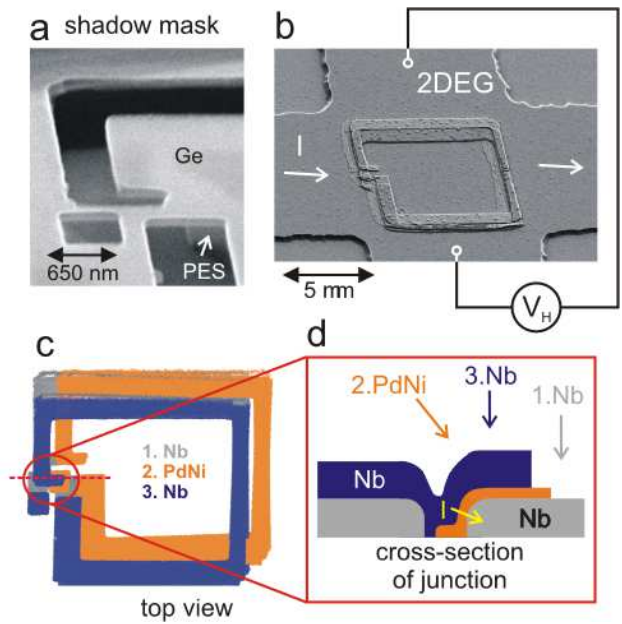}
\caption{\label{fig2} Shadow mask and sample layout. The samples are niobium (Nb) loops with a planar ferromagnetic palladium-nickel ($Pd_{0.82}Ni_{0.18}$) Josephson-junction. \label{fig2a}a) The polyethersulfone (PES)/germanium shadow mask. \label{fig2b}b) Scanning electron micrograph of the sample after lift-off. The loop is placed on top of the flux-sensitive area of the micro-Hall cross made of a GaAs/AlGaAs heterostructure.  \label{fig2c}c) The relative shifts of the three metal films on the substrate are achieved by adjusting different tilt angles before each evaporation. \label{fig2d}d) Cross section of the planar Josephson junction.
}
\end{figure}

The latter experiments concentrated on the high temperature regime, where the induced supercurrents are negligible. In this Letter we focus on the low temperature regime where the $\pi$-junction manifests itself by spontaneous supercurrents, resulting in half integer flux quantization. The spontaneous current induces an asymmetry in the switching fields between different flux states of the loop and it can also be observed directly in the temperature dependent magnetization in absence of an external magnetic flux.

In order to obtain clean S/F-interfaces the different metal layers are evaporated in a single vacuum run using a high-temperature stable shadow mask\cite{Hoss2002,Dubos2000} made of polyethersulfone (PES) and germanium (Ge)(see Fig. \ref{fig2a}a). The samples are niobium square loops with an average line width of 700 nm and a layer thickness of 80 nm. The effective side length $d$ = 7,4 $\mu$m of the $\pi$-loop requires a magnetic field of $B_0$ = 37,5 $\mu$T to obtain one flux quantum in the loop; the average inner diameter of the loop leads to an estimated self-inductance of $L \cong$ 23 pH\cite{Grover1964}. The niobium ring is interrupted by a planar ferromagnetic Josephson junction made of a palladium-nickel alloy ($Pd_{0.82}Ni_{0.18}$)\cite{Kontos2001,Kontos2002} (see Fig. \ref{fig2c}c and Fig. \ref{fig2d}d for details on the geometry). For this nickel concentration the PdNi is a weak ferromagnet\cite{Curie} with a moderate exchange field. The character (0 or $\pi$) of the junction is determined by the ferromagnetic layer thickness; in our case a ferromagnetic layer thickness $d_F$ of 7,5 nm results in a $\pi$-junction ($\xi_F$ = 2,4 nm)\cite{Kontos2002}. For comparison, niobium loops containing a conventional weak tunneling barrier instead of the magnetic interlayer are prepared on the same GaAs chip. The 0-loop has a slightly smaller effective size of d=7,1 $\mu$m corresponding to $B_0$=40,8 $\mu$T.

\begin{figure*}
\includegraphics{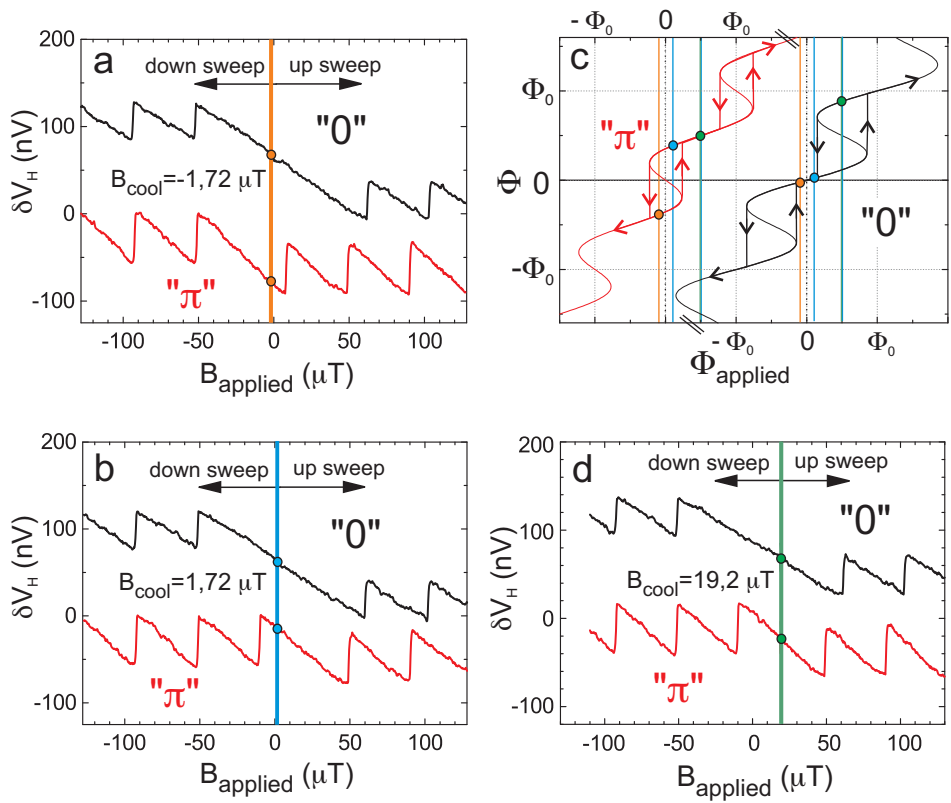}
\caption{\label{fig3} Signature of the $\pi$-state in the switching behavior during magnetic field sweeps.\label{ffig3a}\label{fig3b} a, b) Magnetization traces $\delta V_H \sim I_S$ for a $\pi$- and a 0-loop. All data are taken at 2 K using a measurement current of 20 $\mu$A. The critical currents are 65 $\mu$A for the 0-junction and 27 $\mu$A for the $\pi$-junction. By choosing cooling fields slightly below and above zero applied flux and performing sweeps of the magnetic fields in both field directions, an asymmetric switching behavior of the $\pi$-loop is observed, while the 0-loop shows symmetric jumps with respect to the cooling field.\label{fig3c} c) Relation between the total magnetic flux $\Phi$, applied flux $\Phi_{applied}$ (linear background) and the flux $LI_S$ generated by the supercurrent in the loop. The $\Phi (\Phi_{applied}$) relation of a $\pi$-loop (red) is shifted by $\Phi_0 / 2$ with respect to that of a conventional 0-loop. During the field sweep, only the sections with positive slope can be traced, which leads to a hysteretic switching behavior for $LI_C > \Phi_0 / 2$. The three vertical lines in each curve indicate the applied flux during field cooling below the critical temperature for the three sets of measurements discussed. \label{fig3d} d) Reversal of the symmetry properties of $\pi$- and 0-loop when choosing a cooling field equal to half a flux quantum.}
\end{figure*}

The local magnetic flux of the loop is measured by a micro-Hall sensor in the ballistic limit. Here the Hall voltage $V_H$ is related to the magnetic flux $\Phi$ via $\Phi = \frac{ne}{\alpha} A V_H / I$, where $\alpha = 0.035$ is the geometric filling factor of the loop within the flux sensor, $n$ is the electron density of the sensor, $e$ the electron charge, $I$ the current through the sensor and $A$ the active area of the sensor\cite{Geim1997}. The sensors are made of a modulation doped GaAs/AlGaAs heterostructure containing a two dimensional electron gas 190 nm below the surface. At 4.2 K the heterostructure has a mobility of 750.000 cm/Vs and an electron density of 2.66 *10$^{11}$ cm$^{-2}$.

The magnetic response of a superconducting loop with a conventional Josephson junction is expected to show flux quantization, provided that the width of the superconducting strip is larger than the magnetic penetration depth and that the $LI_C$-product of the loop is larger than $\Phi_0 / 2$\cite{Silver1967,Barone1982}. Figure \ref{fig3} shows the Hall voltage $\delta V_H$ generated by the magnetic flux of the circulating current in the loops at 2K when sweeping the external magnetic field. The contribution of the applied magnetic field has been subtracted. The sample is magnetically shielded by a cryoperm-can, reducing the residual magnetic field to about 4 $\mu$T. Reversing the sweep direction results in a strongly hysteretic magnetization curve (not shown for clarity) and trapped magnetic flux in the loop. To avoid flux trapping, every field sweep plotted in Fig. \ref{fig3} was performed after field cooling down from 10 K, which is higher than the critical temperature of bulk niobium. Each curve contains two measurements in positive and negative sweep direction, respectively. Panels a and b correspond to cooling fields $B_{cool}$= +/- 1,7 $\mu$T, slightly above and below zero magnetic field. The magnetization traces of the 0-loop (black) show diamagnetic screening until the critical current is reached and one flux quantum enters the loop as indicated by a sudden jump of $\delta V_H$. Further rise of $B_{applied}$ is again screened until more flux quanta enter and a $\Phi_0$-periodic sawtooth pattern emerges. At low temperatures, the jump height corresponds to $\Phi_0$ and is used for the determination of the filling factor $\alpha$ for each sample. The magnetic response of the $\pi$-loop looks qualitatively similar but is strongly asymmetric with respect to the applied flux. The origin of the switching behavior is further illustrated in Fig. \ref{fig3c}c, where the black trace shows the multi-valued relation between $\Phi$ and $\Phi_{applied}$ for a hysteretic Josephson junction loop with $LI_C > \Phi_0 / {2\pi}$ \cite{Silver1967,Barone1982}. We again assume a sinusoidal current phase relation\cite{Kulik1978}. The left trace (red) in Fig. \ref{fig3c}c sketches the $\pi$-shifted $\Phi(\Phi_{applied})$ relation expected for a $\pi$-loop. It is seen that the branch near $\Phi=0$ has a negative slope and the corresponding states are unstable. As a consequence the loop should generate a spontaneous current, whose orientation depends on the sign of the small initial cooling field. The essential signature of the $\pi$-shifted $\Phi(\Phi_{applied}$) relation is a pronounced asymmetry of the switching fields, which are defined by the vertical slopes of the red curve in Fig. \ref{fig3c}c. This asymmetry results from the orientation of the spontaneous current selected by the cooling field and is clearly seen in the magnetization curves of the $\pi$-loops in Figs. \ref{fig3}a,b(red). On the other hand, when applying a cool-down field of $B_{cool}$ close to half a flux-quantum (see Fig. \ref{fig3d}d), the situation reverses when compared to $B_{cool} \approx 0$. Now the 0-loop shows asymmetric and the $\pi$-loop shows symmetric switching. This can be understood again from Fig. \ref{fig3c}c: At $\Phi_{applied} = \Phi_0 / 2$ the state of the $\pi$-loop is uniquely defined, whereas for the 0-loop two states compete, which are energetically degenerate.

\begin{figure}
\includegraphics{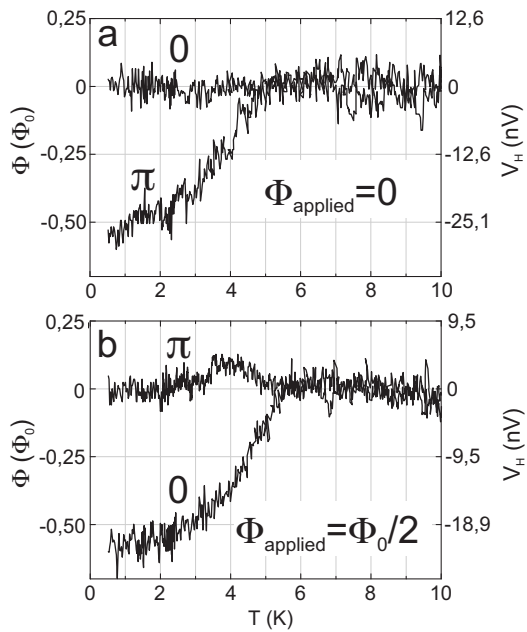}
\caption{\label{fig4} Temperature dependence of the spontaneous current. The measurements show the temperature dependent magnetic flux produced by a $\pi$- and a 0-loop when cooling in \label{fig4a}(a) zero field and in \label{fig4b}(b) a magnetic field equal to half a flux quantum in the loop. Below T $\approx$ 5.5 K the $\pi$-loop develops a spontaneous current when cooling down in zero field, while the magnetic flux through the 0-loop remains zero. When applying a field equal to half a flux quantum $\Phi_0$ the roles of the $\pi$- and 0-loop are exchanged. At low temperatures the spontaneous flux of both loops saturates close to $\Phi_0 / 2$, as expected. To avoid heating of the sensor at the lowest temperatures, the measurement current has been reduced to 7 $\mu$A.}
\end{figure}

Besides the signature of the spontaneous current in the magnetization curves one would like to directly see its emergence when cooling down below $T_C$. Figure \ref{fig4} shows the temperature dependence of the magnetic flux detected in the loops. The raw data of the temperature sweeps contained a large contribution from the temperature dependent longitudinal resistance of the Hall-probe which is nearly independent of the magnetic field\cite{Paalanen1984}. By subtraction of two temperature sweeps at slightly different cooling fields with opposite spontaneous current, the Hall signal has been extracted. The first pair of measurements is taken around $\Phi_{applied}=0$. In order to satisfy the condition of fluxoid quantization the $\pi$-loop is forced to generate a spontaneous circulating current $I_S$. At low temperatures the total flux $\Phi_{total}=LI_S$ approaches $\Phi_0 / 2$. Indeed, the $\pi$-loop shows a small but clearly visible Hall signal as the temperature of the junction is reduced in zero applied field, which results from the magnetic flux generated by the spontaneous current. This can be taken as direct evidence for the \itshape{half integer}\upshape\/ flux quantization induced by the $\pi$-junction. As expected, the 0-loop shows no significant variation of the Hall voltage. Finally, the roles of the $\pi$- and 0-loop can be exchanged by applying an external flux $\Phi_{applied}= \Phi_0 / 2$ during the cool-down. In this case, the Hall voltage of the $\pi$-loop is constant and no spontaneous current arises, as the fluxoid quantization condition is initially compatible with the applied field equivalent to $\Phi_0 / 2$. On the other hand, the 0-loop develops a supercurrent with $LI_S=\Phi_0 / 2$ to provide \itshape{integer}\upshape\/ flux quantization.

In summary, we have demonstrated experimentally the presence of spontaneous magnetic moments in superconducting loops containing a ferromagnetic $\pi$-junction. This is most direct evidence for the existence of inhomogeneous superconductivity in the magnetic layers and can be compared with the direct detection of half flux quanta in tri-crystalline YBCO loops, which are a hallmark of the d-wave symmetry of the order parameter in high-$T_C$ superconductors\cite{Kirtley1995}. The $\pi$-loops can be viewed as a macroscopic analog of a spin 1/2 with two degenerate states at zero magnetic field.

We thank J. Eroms and J. Stahl for technical support and S. Lok for helpful discussions. This work was supported by the Deutsche Forschungsgemeinschaft

\bibliographystyle{apsrev}
\bibliography{SFS_1203}

\begin{thebibliography}{24}
\expandafter\ifx\csname natexlab\endcsname\relax\def\natexlab#1{#1}\fi
\expandafter\ifx\csname bibnamefont\endcsname\relax
  \def\bibnamefont#1{#1}\fi
\expandafter\ifx\csname bibfnamefont\endcsname\relax
  \def\bibfnamefont#1{#1}\fi
\expandafter\ifx\csname citenamefont\endcsname\relax
  \def\citenamefont#1{#1}\fi
\expandafter\ifx\csname url\endcsname\relax
  \def\url#1{\texttt{#1}}\fi
\expandafter\ifx\csname urlprefix\endcsname\relax\def\urlprefix{URL }\fi
\providecommand{\bibinfo}[2]{#2}
\providecommand{\eprint}[2][]{\url{#2}}

\bibitem[{\citenamefont{Josephson}(1962)}]{Josephson1962}
\bibinfo{author}{\bibfnamefont{B.~D.} \bibnamefont{Josephson}},
  \bibinfo{journal}{Phys. Lett.} \textbf{\bibinfo{volume}{1}},
  \bibinfo{pages}{251} (\bibinfo{year}{1962}).

\bibitem[{\citenamefont{Josephson}(1964)}]{Josephson1964}
\bibinfo{author}{\bibfnamefont{B.~D.} \bibnamefont{Josephson}},
  \bibinfo{journal}{Rev. Mod. Phys.} \textbf{\bibinfo{volume}{36}},
  \bibinfo{pages}{216} (\bibinfo{year}{1964}).

\bibitem[{\citenamefont{Kulik and Omelyanchuk}(1978)}]{Kulik1978}
\bibinfo{author}{\bibfnamefont{I.~O.} \bibnamefont{Kulik}} \bibnamefont{and}
  \bibinfo{author}{\bibfnamefont{A.}~\bibnamefont{Omelyanchuk}},
  \bibinfo{journal}{Sov. J. Low Temp. Phys.} \textbf{\bibinfo{volume}{4}},
  \bibinfo{pages}{142} (\bibinfo{year}{1978}), \bibinfo{note}{[Fiz. Nizk. Temp.
  {\bfseries4}, 296 (1978)]}.

\bibitem[{\citenamefont{Heikkil\"a et~al.}(2002)\citenamefont{Heikkil\"a,
  S\"arkk\"a, and Wilhelm}}]{Heikkilae2002}
\bibinfo{author}{\bibfnamefont{T.~T.} \bibnamefont{Heikkil\"a}},
  \bibinfo{author}{\bibfnamefont{J.}~\bibnamefont{S\"arkk\"a}},
  \bibnamefont{and} \bibinfo{author}{\bibfnamefont{F.~K.}
  \bibnamefont{Wilhelm}}, \bibinfo{journal}{Phys. Rev. B}
  \textbf{\bibinfo{volume}{66}}, \bibinfo{pages}{184513}
  (\bibinfo{year}{2002}).

\bibitem[{\citenamefont{Buzdin et~al.}(1982)\citenamefont{Buzdin, Bulaevskii,
  and Panyukov}}]{Buzdin1982}
\bibinfo{author}{\bibfnamefont{A.~I.} \bibnamefont{Buzdin}},
  \bibinfo{author}{\bibfnamefont{L.~N.} \bibnamefont{Bulaevskii}},
  \bibnamefont{and} \bibinfo{author}{\bibfnamefont{S.~V.}
  \bibnamefont{Panyukov}}, \bibinfo{journal}{JETP Lett.}
  \textbf{\bibinfo{volume}{35}}, \bibinfo{pages}{178} (\bibinfo{year}{1982}),
  \bibinfo{note}{[Pis'ma Zh. Eksp. Teor. Fiz. {\bfseries35}, 147]}.

\bibitem[{\citenamefont{Bulaevskii et~al.}(1977)\citenamefont{Bulaevskii,
  Kuzii, and Sobyanin}}]{Bulaevskii1977}
\bibinfo{author}{\bibfnamefont{L.~N.} \bibnamefont{Bulaevskii}},
  \bibinfo{author}{\bibfnamefont{V.~V.} \bibnamefont{Kuzii}}, \bibnamefont{and}
  \bibinfo{author}{\bibfnamefont{A.~A.} \bibnamefont{Sobyanin}},
  \bibinfo{journal}{JETP Lett.} \textbf{\bibinfo{volume}{25}},
  \bibinfo{pages}{290} (\bibinfo{year}{1977}), \bibinfo{note}{[Pis'ma Zh. Eksp.
  Teor. Fiz. {\bfseries25}, 314]}.

\bibitem[{\citenamefont{Chien and Reich}(1999)}]{Chien1999}
\bibinfo{author}{\bibfnamefont{C.~L.} \bibnamefont{Chien}} \bibnamefont{and}
  \bibinfo{author}{\bibfnamefont{D.~H.} \bibnamefont{Reich}},
  \bibinfo{journal}{JMMM} \textbf{\bibinfo{volume}{200}}, \bibinfo{pages}{83}
  (\bibinfo{year}{1999}).

\bibitem[{\citenamefont{Ryazanov et~al.}(2001)}]{Ryazanov2001}
\bibinfo{author}{\bibfnamefont{V.~V.} \bibnamefont{Ryazanov}}
  \bibnamefont{et~al.}, \bibinfo{journal}{Phys. Rev. Lett.}
  \textbf{\bibinfo{volume}{86}}, \bibinfo{pages}{2427} (\bibinfo{year}{2001}).

\bibitem[{\citenamefont{Kontos et~al.}(2001)}]{Kontos2001}
\bibinfo{author}{\bibfnamefont{T.}~\bibnamefont{Kontos}} \bibnamefont{et~al.},
  \bibinfo{journal}{Phys. Rev. Lett.} \textbf{\bibinfo{volume}{86}},
  \bibinfo{pages}{304} (\bibinfo{year}{2001}).

\bibitem[{\citenamefont{Kontos et~al.}(2002)}]{Kontos2002}
\bibinfo{author}{\bibfnamefont{T.}~\bibnamefont{Kontos}} \bibnamefont{et~al.},
  \bibinfo{journal}{Phys. Rev. Lett.} \textbf{\bibinfo{volume}{89}},
  \bibinfo{pages}{137007} (\bibinfo{year}{2002}).

\bibitem[{\citenamefont{Baselmans et~al.}(1999)}]{Baselmans1999}
\bibinfo{author}{\bibfnamefont{J.~J.~A.} \bibnamefont{Baselmans}}
  \bibnamefont{et~al.}, \bibinfo{journal}{Nature}
  \textbf{\bibinfo{volume}{397}}, \bibinfo{pages}{43} (\bibinfo{year}{1999}).

\bibitem[{\citenamefont{Baselmans
  et~al.}(2002{\natexlab{a}})\citenamefont{Baselmans, van Wees, and
  Klapwijk}}]{Baselmans2002_1}
\bibinfo{author}{\bibfnamefont{J.~A.} \bibnamefont{Baselmans}},
  \bibinfo{author}{\bibfnamefont{B.~J.} \bibnamefont{van Wees}},
  \bibnamefont{and} \bibinfo{author}{\bibfnamefont{T.~M.}
  \bibnamefont{Klapwijk}}, \bibinfo{journal}{Phys. Rev. B}
  \textbf{\bibinfo{volume}{65}}, \bibinfo{pages}{224513}
  (\bibinfo{year}{2002}{\natexlab{a}}).

\bibitem[{\citenamefont{Baselmans
  et~al.}(2002{\natexlab{b}})}]{Baselmans2002_2}
\bibinfo{author}{\bibfnamefont{J.~J.~A.} \bibnamefont{Baselmans}}
  \bibnamefont{et~al.}, \bibinfo{journal}{Phys. Rev. Lett.}
  \textbf{\bibinfo{volume}{89}}, \bibinfo{pages}{107002}
  (\bibinfo{year}{2002}{\natexlab{b}}).

\bibitem[{\citenamefont{Kirtley et~al.}(1995)}]{Kirtley1995}
\bibinfo{author}{\bibfnamefont{J.~R.} \bibnamefont{Kirtley}}
  \bibnamefont{et~al.}, \bibinfo{journal}{Nature}
  \textbf{\bibinfo{volume}{373}}, \bibinfo{pages}{225} (\bibinfo{year}{1995}).

\bibitem[{\citenamefont{Guichard et~al.}(2003)}]{Guichard2003}
\bibinfo{author}{\bibfnamefont{W.}~\bibnamefont{Guichard}}
  \bibnamefont{et~al.}, \bibinfo{journal}{Phys. Rev. Lett.}
  \textbf{\bibinfo{volume}{90}}, \bibinfo{pages}{167001}
  (\bibinfo{year}{2003}).

\bibitem[{\citenamefont{Ryazanov et~al.}(2002)}]{Ryazanov2002}
\bibinfo{author}{\bibfnamefont{V.~V.} \bibnamefont{Ryazanov}}
  \bibnamefont{et~al.}, \bibinfo{journal}{Phys. Rev. B}
  \textbf{\bibinfo{volume}{65}}, \bibinfo{pages}{020501}
  (\bibinfo{year}{2002}).

\bibitem[{\citenamefont{Hoss et~al.}(2002)}]{Hoss2002}
\bibinfo{author}{\bibfnamefont{T.}~\bibnamefont{Hoss}} \bibnamefont{et~al.},
  \bibinfo{journal}{Physica E} \textbf{\bibinfo{volume}{14}},
  \bibinfo{pages}{341} (\bibinfo{year}{2002}).

\bibitem[{\citenamefont{Dubos et~al.}(2000)}]{Dubos2000}
\bibinfo{author}{\bibfnamefont{P.}~\bibnamefont{Dubos}} \bibnamefont{et~al.},
  \bibinfo{journal}{J. of Vac. Sci. and Technol. B}
  \textbf{\bibinfo{volume}{18}}, \bibinfo{pages}{122} (\bibinfo{year}{2000}).

\bibitem[{\citenamefont{Grover}(1962)}]{Grover1964}
\bibinfo{author}{\bibfnamefont{F.~W.} \bibnamefont{Grover}},
  \emph{\bibinfo{title}{Inductance Calculations: Working Formulas and Tables}}
  (\bibinfo{publisher}{Dover Publ.}, \bibinfo{address}{New York},
  \bibinfo{year}{1962}).

\bibitem[{Cur()}]{Curie}
\bibinfo{note}{The Curie temperature measured by anomalous Hall effect in a
  bare thin film is 200K}.

\bibitem[{\citenamefont{Geim et~al.}(1997)}]{Geim1997}
\bibinfo{author}{\bibfnamefont{A.~K.} \bibnamefont{Geim}} \bibnamefont{et~al.},
  \bibinfo{journal}{Appl. Phys. Lett.} \textbf{\bibinfo{volume}{71}},
  \bibinfo{pages}{2379} (\bibinfo{year}{1997}).

\bibitem[{\citenamefont{Silver and Zimmerman}(1967)}]{Silver1967}
\bibinfo{author}{\bibfnamefont{A.~H.} \bibnamefont{Silver}} \bibnamefont{and}
  \bibinfo{author}{\bibfnamefont{J.~E.} \bibnamefont{Zimmerman}},
  \bibinfo{journal}{Phys. Rev.} \textbf{\bibinfo{volume}{157}},
  \bibinfo{pages}{317} (\bibinfo{year}{1967}).

\bibitem[{\citenamefont{Barone and Paterno}(1982)}]{Barone1982}
\bibinfo{author}{\bibfnamefont{A.}~\bibnamefont{Barone}} \bibnamefont{and}
  \bibinfo{author}{\bibfnamefont{G.}~\bibnamefont{Paterno}},
  \emph{\bibinfo{title}{Physics and applications of the Josephson effect}}
  (\bibinfo{publisher}{Wiley}, \bibinfo{address}{New York},
  \bibinfo{year}{1982}).

\bibitem[{\citenamefont{Paalanen et~al.}(1984)}]{Paalanen1984}
\bibinfo{author}{\bibfnamefont{M.~A.} \bibnamefont{Paalanen}}
  \bibnamefont{et~al.}, \bibinfo{journal}{Phys. Rev. B}
  \textbf{\bibinfo{volume}{29}}, \bibinfo{pages}{6003} (\bibinfo{year}{1984}).

\end{thebibliography}

\end{document}